\documentclass[letterpaper, 10 pt, conference]{ieeeconf}
\IEEEoverridecommandlockouts
\DeclareMathAlphabet{\mbf}{OT1}{ptm}{b}{n}
\newcommand{\trans}{{\ensuremath{\mathsf{T}}}}
\usepackage[backend=biber,
            bibstyle=ieee,
            citestyle=numeric,
            sorting=none,
            sortcites=true]{biblatex}

\usepackage[margin=0.75in]{geometry}
\usepackage{amsmath}
\usepackage{graphicx}
\usepackage{graphics}
\usepackage{subcaption}
\usepackage{tikz, circuitikz, siunitx}
\usepackage{blindtext}

\title{\LARGE \bf Power Hardware-in-the-loop Interfacing via
$\mathcal{H}_\infty$ Model Matching}

\author{Jonathan Eid$^{1}$, Ashley Meagher$^{1}$, Dmitry Rimorov$^{2}$,\\ Anil
Kumar Bonala$^{3}$, Rajendra Thike$^{3}$, and James Richard Forbes$^{1}$
\thanks{This work was supported by the NSERC Alliance program, Hydro-Qu\'{e}bec,
and OPAL-RT Technologies.}
\thanks{$^{1}$J.~Eid, A.~Meagher, and J.R.~Forbes are with the Dept.~of
Mechanical Engineering, McGill University. e-mail: \texttt{ \{jonathan.eid,
ashley.meagher\}@mail.mcgill.ca, james.richard.forbes@mcgill.ca}}
\thanks{$^{2}$D.~Rimorov is with the Dept.~of Power Electronics in Grids,
Hydro-Qu\'{e}bec Research Institute. e-mail:
\texttt{rimorov.dmitry@hydroquebec.com}}
\thanks{$^{3}$A.K.~Bonala and R.~Thike are with the Div.~of Power Applications
and Solutions, OPAL-RT Technologies. e-mail: \texttt{\{anil.kumarbonala,
rajendra.thike\}@opal-rt.com}}
}

\begin{document}

\maketitle

% Abstract
\begin{abstract}

This paper presents an $\mathcal{H}_\infty$ model matching control approach to
the problem of power hardware-in-the-loop (PHIL) interfacing.  The objective is
to interconnect a grid simulation and a physical device via an interface in a
way that is stable and accurate.  Conventional approaches include the ideal
transformer method (ITM) and its impedance-based variants, which trade accuracy
for stability, as well as some $\mathcal{H}_\infty$ control-based approaches,
which do not make use of all the available information in their optimization for
accuracy.  Designing for transparency, as opposed to accuracy as existing
approaches do, would achieve both stability and accuracy, while making use of
all the dynamical information present in the idealized interconnection of the
grid and device.  The approach proposed in this paper employs model matching to
formulate the PHIL problem as an $\mathcal{H}_\infty$ control problem using
transparency as the explicit frequency-domain control objective. The approach is
experimentally validated in a real-time resistive-load PHIL setup, and is found
to achieve accuracy levels that are comparable or superior to those of an
ITM-based interface.

\end{abstract}


\section{Introduction}\label{sec:1}

To combat climate change, the development and utilization of sustainable energy
sources has become increasingly critical.  As the pace of the sustainable energy
transition accelerates, power system operators are faced with the challenge of
integrating a diverse set of power generation technologies into the grid,
particularly renewable energy sources (RES)~\cite{bhattarai}.  To
understand how RES impact the power grid, testing is essential. However,
full-scale in-field test integration of RES is often costly and impractical.
Safe, reliable, and affordable integration testing represents a crucial step in
enabling the ongoing energy transition.  Power hardware-in-the-loop (PHIL)
infrastructure provides a laboratory-based real-time testing platform, and has
become vital for evaluating system performance under realistic
conditions~\cite{vonjouanne}.

\begin{figure}[t]
    \centering
    \begin{subfigure}[b]{\columnwidth}
        \input{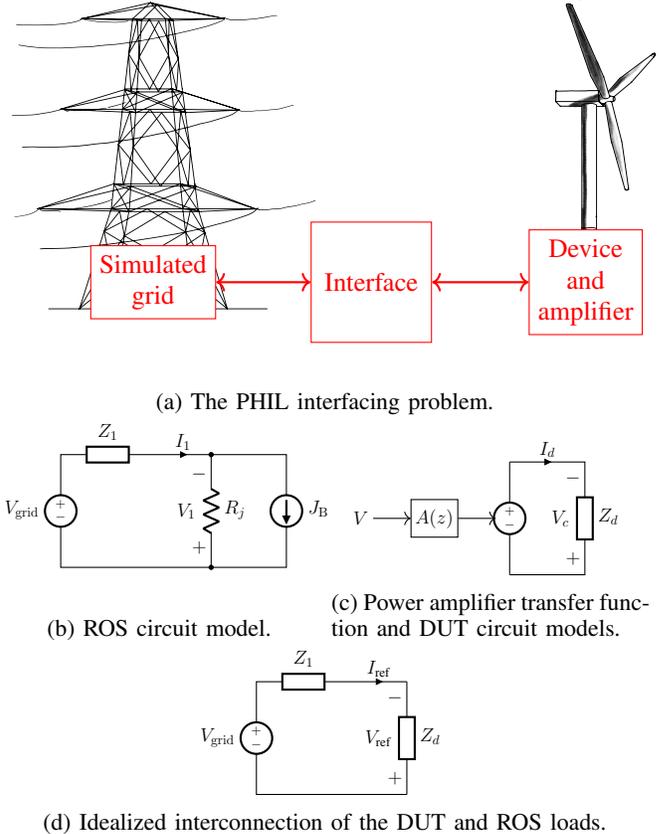}
        \caption{The PHIL interfacing problem.}
        \label{fig:1_a}
    \end{subfigure}
    \begin{subfigure}[b]{0.49\columnwidth}
        \centering
        \begin{circuitikz}[
    scale=0.5,
    font=\Large,
    american,
    transform shape]
    \draw (0,0)
        to [short, *-] (-4,0)
        to [V, l=$V_\mathrm{grid}$, invert] (-4,3)
        to [generic, l=$Z_1$] (-1.5,3)
        to [short, i=$I_1$, -*] (0,3)
        ;
    \draw (0,3)
        to [R, l=$R_j$, v<=$V_1$] (0,0)
        ;
    \draw (0,3)
        to [short] (2,3)
        to [I, l=$J_\mathrm{B}$] (2,0)
        to [short] (0,0)
        ;
\end{circuitikz}
        \caption{ROS circuit model.}
        \label{fig:1_b}
    \end{subfigure}\hfill
    \begin{subfigure}[b]{0.49\columnwidth}
        \centering
        \begin{circuitikz}[
    scale=0.5,
    font=\Large,
    american,
    transform shape]
    \node [draw, rectangle, minimum size=10 mm] (amp) at (11 - 2, 1.5) {$A(z)$};
    \draw [<-] (amp.west) -- ++(-1, 0) node[anchor=east] {$V$};
    \draw [->] (amp.east) -- ++(1, 0);
    \draw (11,0)
        to [V, invert] (11,3)
        ;
    \draw (11,3)
        to [short, i=$I_d$] (13,3)
        to [generic, l=$Z_d$, v<=$V_c$] (13,0)
        to [short] (11,0)
        ;
\end{circuitikz}
        \caption{Power amplifier transfer function and DUT circuit models.}
        \label{fig:1_c}
    \end{subfigure}
    \begin{subfigure}[b]{\columnwidth}
        \centering
        \begin{circuitikz}[
    scale=0.5,
    font=\Large,
    american,
    transform shape]
    \draw (0,0)
        to [short] (-4,0)
        to [V, l=$V_\mathrm{grid}$, invert] (-4,3)
        to [generic, l=$Z_1$] (-1.5,3)
        to [short, i=$I_\text{ref}$] (0,3)
        ;
    \draw (0,3)
        to [generic, l=$Z_d$, v<=$V_\text{ref}$] (0,0)
        ;
\end{circuitikz}
        \caption{Idealized interconnection of the DUT and ROS loads.}
        \label{fig:3_a}
    \end{subfigure}
    \caption{PHIL interfacing problem and its components.}
    \label{fig:1}
    \vskip -6mm
\end{figure}

PHIL simulations consist of four key components. They are the physical device
under test (DUT), the power amplification unit, the interfacing algorithm, and
the rest of the system (ROS), which is a real-time electrical grid
simulation~\cite{lauss}. Figures~\ref{fig:1_b}~and~\ref{fig:1_c}
depict the circuit and transfer function models of the ROS, DUT, and amplifier,
as well as their voltage and current signals.

A key characteristic of a successful PHIL simulation is its
stability~\cite{ren}. In particular, the ROS and DUT
voltage and current signals must not diverge.
Equally important is performance. In the context of PHIL interfacing, the
two main notions of performance are accuracy and transparency. For
accuracy~\cite{ren}, the voltages and currents of the ROS and DUT must converge
to each other in steady state. For transparency~\cite{naerum}, the voltages and
currents must converge to those of an idealized interconnection of the grid and
device in steady state, as shown in Figure~\ref{fig:3_a}. Following industrial
and research convention, accuracy will be used throughout this paper as the
performance metric for experimental validation and comparison.
The problem under consideration, depicted in Figure~\ref{fig:1_a}, is the design
of a PHIL interfacing algorithm that ensures both performance and stability.

Several approaches have been proposed for the design of PHIL
interfaces~\cite{ren,lauss,resch}.
One of the most common is the ideal transformer method (ITM), which can be
implemented as either a voltage-controlled or a current-controlled variant.
Referring to Figures~\ref{fig:1_b}~and~\ref{fig:1_c}, a voltage-controlled
ITM-based interface provides the ROS voltage $V_1$ to the DUT amplifier as its
voltage command $V$, which produces the DUT current $I_d$. That current is then
measured, filtered, and fed back to the ROS as its actuation $J_B$.
Voltage- or current-controlled interface alternatives such as the partial
circuit duplication (PCD) or damping impedance methods (DIM) attempt to modify
the ITM-based interface by introducing artificial impedance elements in the ROS
and DUT circuits. This leads to improvements in stability by the PCD, and
improvements in accuracy by the DIM, given specific knowledge of the ROS and DUT
dynamics.

The transmission line method (TLM)~\cite{tremblay2} combines voltage and current
measurements from the DUT and ROS as well as an artificial impedance to generate
propagating wave signals, modeled after those of a transmission line, that are
used as the actuation commands of the ROS and DUT. This leads to strong
stability guarantees, and some degree of accuracy in specific configurations.

However, all of these voltage- and current-controlled methods do not make use of
a systematic design process, making it difficult to achieve both the required
stability and accuracy. On the other hand, there are generalized control methods
that apply control-theoretic tools to design the interface, using
$\mathcal{H}_\infty$ control~\cite{salapaka1,salapaka2} to systematically
achieve stability and accuracy.
These control-theoretic methods start from clear performance specifications, and
design the interface to meet them, using formal optimization theory and mature
optimization tools to allow for greater reliability, and easier adaptation to
different parameters.
Additionally, the foundational theory of \(\mathcal{H}_\infty\)-optimal control
is uniquely suitable to address the PHIL interfacing performance problem, as
will be explained in Section~\ref{sec:2}, as well as address the PHIL
interfacing robustness problem, via the tools of \(\mu\) analysis and synthesis
\cite{salapaka2,scherer}.
However, existing \(\mathcal{H}_\infty\) control-based approaches to the PHIL
interfacing problem do not make use of all the available information. This is
because they choose accuracy as their control objective, as opposed to
transparency. By choosing transparency as the control objective, an interfacing
approach would make use of the dynamical information present in the direct
interconnection of the grid and device. It would also make the stronger
performance guarantee of transparency, which implies accuracy.

The model matching control architecture~\cite{scherer} provides a
systematic way to incorporate the objective of transparency into the design of
PHIL interfaces. This makes use of all the information present in the idealized
interconnection of the grid and device, and achieves accuracy via transparency.

The novel contribution of this paper is an $\mathcal{H}_\infty$
model matching control approach to the design of PHIL interfaces that allows for the
systematic encoding of transparency as the control objective. The approach is
experimentally validated on a real-time PHIL experiment. Both interface
stability and accuracy are confirmed experimentally. In particular, the accuracy
of the proposed approach is shown to be comparable or better than that of the
ITM-based method in experiments.

The rest of the paper is organized as follows.
Section~\ref{sec:2} presents the mathematics necessary to formulate the PHIL
interfacing problem as a model matching problem. Section~\ref{sec:3} outlines
the methodology proposed to obtain a model matching-based PHIL interface.
Section~\ref{sec:4} presents a validation experiment of the methodology, delving
into further details concerning the modeling and time-domain response analysis
of the ROS and DUT. Finally, Section~\ref{sec:5} provides concluding remarks on
the results of the experiment.
% Section~\ref{sec:2} outlines the PHIL
% interfacing problem and how it is formed as a model matching generalized control
% problem. Section III outlines the proposed methodology of the interface design.
% Section IV presents the results of the experimental validation. Concluding
% remarks are provided in Section V.

\section{Preliminaries}\label{sec:2}

\subsection{Notation}

Throughout the paper, $s$ and $z$ respectively denote the Laplace domain and the
$z$ domain. Bold, capital latin letters, such as $\mbf{P}$, denote multi-input,
multi-output transfer matrices. Nonbold latin letters, such as $R_1$, denote
one-dimensional quantities such as constants, signals, and transfer functions of
a single input and output.

\subsection{The generalized control problem}

\begin{figure}
    \centering
    \begin{tikzpicture}
    \node [
        draw,
        rectangle,
        minimum size=10mm,
    ] (P) at (0,0) {$\mbf{P}$};

    \node [
        draw,
        rectangle,
        dashed,
        minimum size=10mm,
    ] (K) at (0,-1.2) {$\mbf{K}$};

    \draw[->] ($(P.west)+(-1,0.2)$) 
            node [left] {$\mbf{w}$}
        -- ($(P.west)+(0,0.2)$)
        ;
    \draw[->] ($(P.west)+(-1,-0.2)$) 
            node [left] {$\mbf{u}$}
        -- ($(P.west)+(0,-0.2)$)
        ;
    \draw[->] ($(P.east)+(0,0.2)$)
        -- ++(1,0)
            node [right] {$\mbf{z}$}
        ;
    \draw[->] ($(P.east)+(0,-0.2)$)
        -- ++(1,0)
            node [right] {$\mbf{y}$}
        ;
    
    \draw[->] (K.west)
        -- ++(-1,0)
            node [left] {$\mbf{u}$}
        ;
    \draw[->] ($(K.east)+(1,0)$)
            node [right] {$\mbf{y}$}
        -- (K.east)
        ;
\end{tikzpicture}
    \caption{Generalized plant and controller.}
    \label{fig:2}
\end{figure}
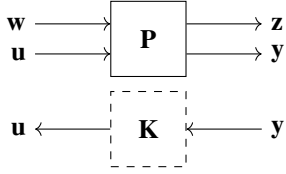

The generalized plant, illustrated in Figure~\ref{fig:2}, is a foundational tool
used in this paper~\cite{skogestad,zhou_doyle_glover}.  It is a general
framework for modeling physical plants as linear, time-invariant (LTI) dynamical
systems. It describes the interaction between the exogenous inputs $\mbf{w}$,
actuation $\mbf{u}$, control input $\mbf{y}$, and the performance variable
$\mbf{z}$, as well as any normalization or frequency-domain filtering, as
discussed in Section~\ref{sec:3}. The generalized plant $\mbf{P}$ is needed for
$\mathcal{H}_\infty$ controller design.

In particular, the $\mathcal{H}_\infty$ controller design procedure synthesizes a
controller $\mbf{K}$ that ensures the asymptotic stability of the closed loop,
which is denoted $\mathcal{F}_\ell(\mbf{P}, \mbf{K})$~\cite[\S3.8]{skogestad},
and that minimizes the gain from the exogenous inputs to the performance
variable outputs. For discrete-time control, the $\mathcal{H}_\infty$ controller
$\mbf{K}$ places the poles of $\mathcal{F}_\ell(\mbf{P}, \mbf{K})$ inside the
unit disk and minimizes the $\mathcal{H}_\infty$ norm
\begin{equation*}
    \|\mathcal{F}_\ell(\mbf{P}, \mbf{K})\|_\infty
    = \sup_{\omega \in [-\pi, \pi]} \bar{\sigma}(\mathcal{F}_\ell(\mbf{P},
    \mbf{K}) (e^{j \omega})),
\end{equation*}
where $\bar{\sigma}$ and $\mathrm{sup}$ denote the maximum singular value and
the supremum, respectively.

Once the generalized plant is computed, a number of well-established synthesis
protocols can be used to generate an appropriate controller, including those
based upon algebraic Riccati equations~\cite{zhou_doyle_glover} or linear matrix
inequality-constrained semidefinite
programs~\cite{deoliveira_geromel_bernussou,caverly_forbes}, for example.

\subsection{The PHIL interfacing model}

The purpose of this section is to explain how the key components of the power
hardware-in-the-loop interfacing problem are modeled within the generalized
plant. Starting from Figures~\ref{fig:1_b}~and~\ref{fig:1_c}, continuous-time
transfer matrix models for the ROS and DUT circuit dynamics are derived using
Kirchhoff's and Ohm's circuit laws. For the ROS circuit, this yields
\begin{align}
    \hspace{-7pt}
    \begin{bmatrix}
        V_1 (s)\\I_1(s)
    \end{bmatrix} &= \mbf{G}_\text{ROS}(s) \begin{bmatrix}
        V_\text{grid}(s)\\J_B(s)
    \end{bmatrix} \\
    &= \frac{1}{R_j + Z_1(s)} \begin{bmatrix}
            R_j && -R_j Z_1(s)\\1 && R_j
        \end{bmatrix} \begin{bmatrix}
            V_\text{grid}(s)\\J_B(s)
        \end{bmatrix},
\end{align}
where $R_j$ and $Z_1(s)$ are the current source's shunt resistor and the
circuit's Thévenin equivalent impedance, respectively. For the DUT and
amplifier, this yields
\begin{align}
    \begin{bmatrix}
        V_c(s)\\I_d(s)
    \end{bmatrix} &= \mbf{G}_\text{DUT}(s) V(s)
    = A(s) \begin{bmatrix}
        1\\Z_d(s)^{-1}
    \end{bmatrix} V(s),
\end{align}
where $A(s)$ is a transfer function model of the amplifier obtained
experimentally by system identification, and $Z_d(s)$ is the DUT load.

These continuous-time models must be \textit{discretized} appropriately. The ROS
being digitally simulated in real time, its circuit is discretized using the
bilinear transform~\cite{tremblay1} to preserve the frequency-domain
response. The power amplifier transfer function and DUT circuit are discretized
using the zero-order hold method to reflect the fact that they are physical
devices whose measurements are sampled in real time. The discrete-time models of
the key components are then used as the building blocks of the generalized plant
of the control problem.

\subsection{Transparency and model matching}

The purpose of this section is to describe the objective of the PHIL interfacing
problem. The exogenous signal of the present version of the PHIL problem is the
grid voltage, $\mbf{w}(z) = V_\text{grid}(z)$. The generalized plant $\mbf{P}(z)$
amplifies the frequency-dynamics of $\mbf{w}(z)$ to yield the performance
variable $\mbf{z}(z)$.
It is the role of the $\mathcal{H}_\infty$-optimal
controller to attenuate this amplification as much as possible, and so
$\mbf{z}(z)$ is chosen to include signals whose frequency-domain gain and
time-domain magnitude must be minimized. The standard choice of $\mbf{z}(z)$
consists of accuracy errors and actuator commands, which in the present context
is
\begin{align}
    \label{eqn:standard_objective}
    \mbf{z}(z) = \begin{bmatrix}
        V_1(z) - V_c(z)\\
        I_1(z) - I_d(z)\\
        V(z)\\
        J_B(z)
    \end{bmatrix}.
\end{align}

It is the primary contribution of this paper to reformulate this PHIL
interfacing problem in a way that is more general. Firstly, a closer look must
be taken at the difference between an \textit{accurate}~\cite{ren} and
\textit{transparent}~\cite{naerum} PHIL interface. An \textit{accurate} PHIL
interface is one where the DUT voltage and current respectively converge to the
ROS voltage and current in steady state. This, in addition to regulating the
actuation effort, is the explicit choice made in the definition of $\mbf{z}(z)$
in (\ref{eqn:standard_objective}).  Subtly distinct is a \textit{transparent}
PHIL interface, which is one where the DUT and ROS voltages and currents
converge to those of the direct interconnection of the DUT and ROS loads in
steady state. This direct interconnection, illustrated in Figure~\ref{fig:3_a},
is free of delays, power amplifiers, and actuation channels, and represents an
idealization. For the purpose of generalized control, this direct
interconnection is modeled in continuous time using Kirchhoff's and Ohm's laws by
the transfer matrix
\begin{align}
    \begin{bmatrix}
        V_\text{ref} (s)\\I_\text{ref}(s)
    \end{bmatrix} &= \mbf{G}_\text{REF}(s) V_\text{grid}(s) \\
    &= \frac{1}{Z_d(s) + Z_1(s)} \begin{bmatrix}
            Z_d(s) \\ 1
        \end{bmatrix} V_\text{grid}(s),
\end{align}
and discretized using the bilinear transform.

\begin{figure}[t!]
    \centering
    \begin{tikzpicture}[scale=0.75]
    % Nodes for the blocks
    \node[draw, minimum size=8mm, align=center] (Pr) at (0, 0) {$\mbf{P}_r$};
    \node[draw, minimum size=8mm, align=center] (Pa) at (0, -12mm) {$\mbf{P}_c$};

    % Input arrows for P_r
    \draw[<-] (Pr.west) -- ++(-10mm, 0) node[anchor=east] {$\mbf{w}$};

    % Input arrows for P_a
    \draw[<-] ($(Pa.west) + (0, 1.667mm)$) -- ++(-10mm, 0) node[anchor=east] {$\mbf{w}$};
    \draw[<-] ($(Pa.west) + (0, -1.667mm)$) -- ++(-10mm, 0) node[anchor=east] {$\mbf{u}$};

    % Outputs for P_r
    \node[draw, circle, inner sep=0pt, minimum size=3mm] (sumV) at (16.5mm,0) {$+$};
    \draw[->] (Pr.east) -- node[midway, above] {$\mbf{y}_r$} (sumV.west);
    \draw[->] (sumV.east) -- ++(10mm, 0) node[right] {$\mbf{e}$};

    % Outputs for P_a
    % \draw[<-] (sumV.south) -- ++(0,-10.5mm) node[right] {$-1$} -- node[midway, above] {$\mbf{y}_a$} (Pa.east);
    \draw[->] (Pa.east) 
        -- node[midway, above] {$\mbf{y}_c$} ($(sumV)+(0,-12mm)$) 
        -- node[pos=0, right] {$-1$} (sumV.south);
    % \draw[->] (Pa.east) -- ++(10.5mm, 0) node[midway, above] {$\mbf{y}_a$} -- ++(1.5mm,0) node[right] {$-1$} -- (sumV.south);

\end{tikzpicture}
    \caption{Generalized plant of the model matching control architecture, where
    $\mbf{z}(t) := [\mbf{e}(t)\; \mbf{u}(t)]^\trans$ and $\mbf{y}(t) :=
    \mbf{y}_c(t)$.}
    \label{fig:3_b}
\end{figure}
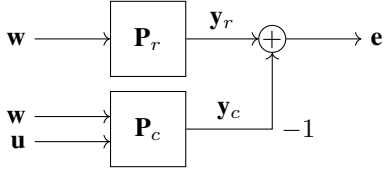

Secondly, the standard PHIL control objective is generalized by employing a
model matching control architecture~\cite{scherer}, illustrated
in Figure~\ref{fig:3_b}. It is designed to make the dynamics of a controlled
plant, $\mbf{P}_c(z)$, \textit{match} those of a reference plant,
$\mbf{P}_r(z)$, as well as regulate the actuation effort. The generalization
comes from the fact that the PHIL interface can be systematically designed to
match any interface that can be modeled as a circuit or transfer matrix.

The design of a transparent PHIL interface proposed in this paper begins by
specializing the general model matching architecture, using
$\mbf{G}_\text{DUT}(z)$, $\mbf{G}_\text{ROS}(z)$, and $\mbf{G}_\text{REF}(z)$ to
select $\mbf{P}_c(z)$ and $\mbf{P}_r(z)$ such that
\begin{align}
    \label{eqn:transparency_objective}
    \mbf{z}(z) = \begin{bmatrix}
        V_1(z)-V_\text{ref}(z)\\
        I_1(z)-I_\text{ref}(z)\\
        V_c(z)-V_\text{ref}(z)\\
        I_d(z)-I_\text{ref}(z)\\
        V(z)\\
        J_B(z)
    \end{bmatrix}
\end{align}
becomes the performance variable.

A crucial remark must be made about transparent interfaces. Given that $V_1(t)$
and $V_c(t)$ will both converge to $V_\text{ref}(t)$, it is clear that $V_1(t) -
V_c(t)$ will converge to $0$. Similarly, $I_1(t)$ and $I_d(t)$ will both
converge to $I_\text{ref}(t)$, making $I_1(t) - I_d(t)$ converge to $0$. Thus, a
transparent interface is also an accurate one.

\section{Methodology}\label{sec:3}

The proposed interfacing approach follows a structured sequence that begins with
deriving first-principle models of the ROS, DUT, REF circuits, as is done in
Section~II.

The PHIL interfacing problem is then formulated as a special case of the model
matching problem, enabling the explicit treatment of transparency as a design
objective. 

Once the model matching generalized plant is defined, an appropriate delay
structure is identified and imposed on the measurement and actuation channels,
$\mbf{y}(z)$ and $\mbf{u}(z)$, to capture the latency characteristics inherent
to the PHIL experiment setup.

Each channel of the generalized plant is then normalized by its maximum expected
time-domain magnitude to render it unitless, specifying the desired performance
and simplifying the loop-shaping process.

The exogenous input and performance channels, $\mbf{w}(z)$ and $\mbf{z}(z)$, are
then augmented with filters to shape the closed-loop response in accordance with
the desired performance specifications. Working with a normalized generalized
plant, this amounts to selecting a set of filters to encode the frequency
content of $\mbf{w}(z)$ and to specify the frequency bandwidths over which each
element of $\mbf{z}(z)$ must be minimized.

Next, the controller is synthesized in discrete time, and the frequency-domain
response of the closed loop is examined to check the performance guarantees.
These guarantees meet the specifications when the gain response of each
input-output channel of the closed loop is less than $0$~dB in the desired
frequency bandwidths.

Finally, the controller is implemented in the real-time PHIL setup to validate
its performance under experimental conditions.

\section{Experimental Validation}\label{sec:4}

\begin{figure}[t]
    \centering
    \begin{tikzpicture}

  % --- Background image ---
  \node[anchor=south west, inner sep=0] (img) at (0,0)
    {\includegraphics[height=0.5\columnwidth]{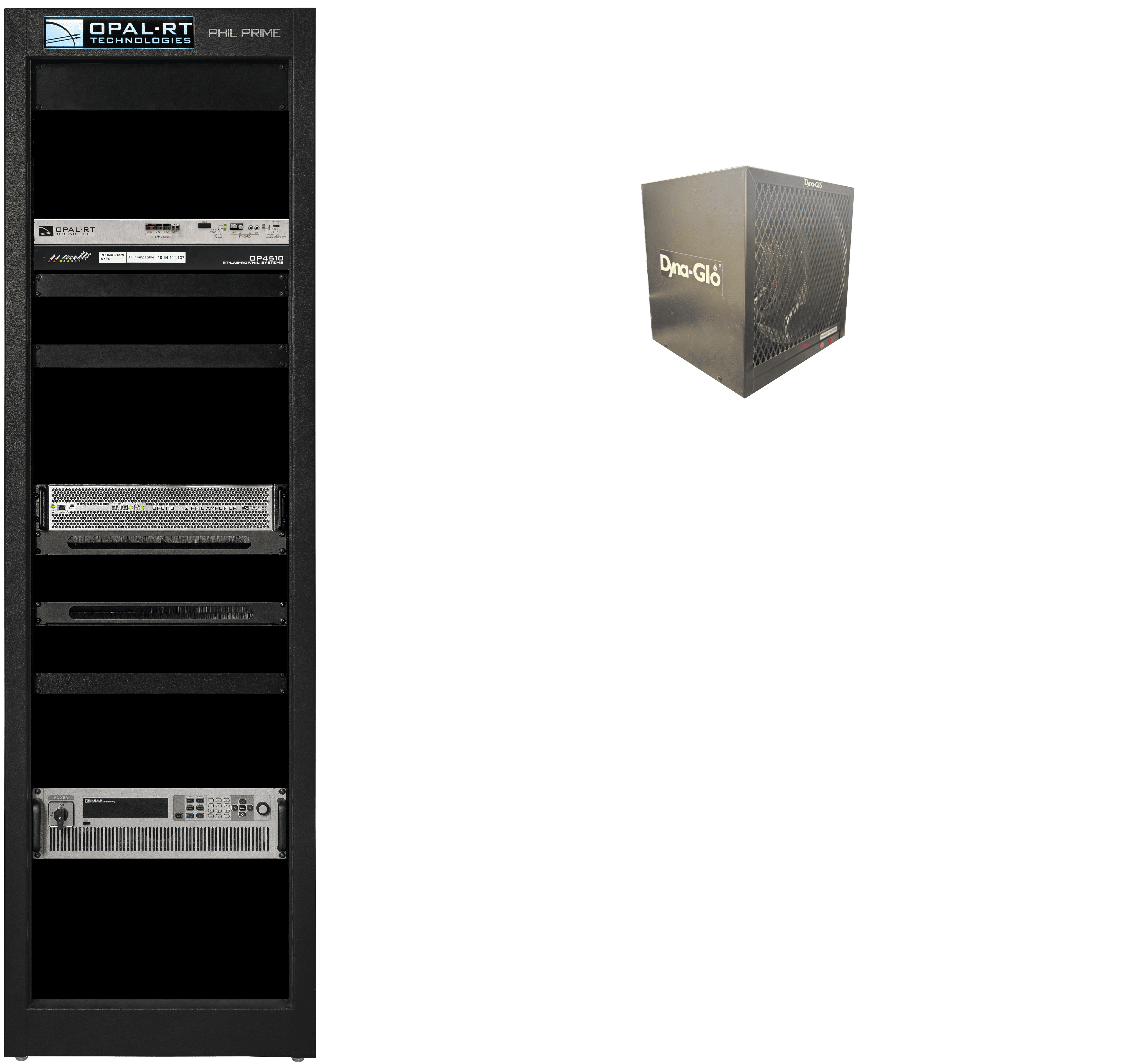}};

  % Use the image coordinate system
  \begin{scope}[
    x={(img.south east)},
    y={(img.north west)}
  ]

  % --- Label boxes ---
  % Simulator (top left, OPAL-RT rack top section)
  \node[draw, fill=white, fill opacity=0.85, text opacity=1,
        rounded corners=2pt, font=\small, anchor=west]
    at (-0.28, 0.70) {Simulator};

  % Power amplifier (left center, large unit mid-rack)
  \node[draw, fill=white, fill opacity=0.85, text opacity=1,
        rounded corners=2pt, font=\small, anchor=west]
    at (-0.45, 0.44) {Power amplifier};

  % DC source (left bottom)
  \node[draw, fill=white, fill opacity=0.85, text opacity=1,
    rounded corners=2pt, font=\small, anchor=west]
    at (-0.31, 0.16) {DC source};

  % Heater (right, Dyna-Glo unit)
  \node[draw, fill=white, fill opacity=0.85, text opacity=1,
        rounded corners=2pt, font=\small, anchor=west]
    at (0.72, 0.70) {Heater, \(R\)};

  % --- Three-phase resistor circuit (below heater) ---
  % Each phase: wire -- R -- R -- wire, stacked vertically
  \ctikzset{american, transform shape}
  \draw (0.65, 0.48) -- (0.93, 0.48) -- (0.93, 0.42)
        to[R, l=\scriptsize$R$, resistors/scale=0.4] (0.93, 0.32)
        to[R, l=\scriptsize$R$, resistors/scale=0.4] (0.93, 0.12) -- (0.65,0.12)
        to[V, invert, l=$V_\text{ABC}$] (0.65,0.48)
        ;
  \draw (0.88, 0.44) rectangle (1.03, 0.16);
  \node[anchor=west] at (1.03, 0.30) {$Z_d$};

  \end{scope}
\end{tikzpicture}
    \caption{PHIL experimental setup, where \(V_\text{ABC}\) is the three-phase voltage generated by the power amplifier.}
    \label{fig:5_a_b}
\end{figure}

The purpose of this section is to report on an experimental application of the
proposed methodology that demonstrates the viability of the model matching-based
approach to PHIL interfacing.

As part of the setup for the experimental validation, the ROS is modeled in
\textit{Simulink} as the Thévenin equivalent circuit depicted in
Figure~\ref{fig:1_b}, built for simulation with \textit{RT-LAB}, and ran on an
\textit{OP4510} simulator, shown in Figure~\ref{fig:5_a_b}. The exogenous signal
$V_\text{grid}$ is taken to be a three-phase sinusoid with a root-mean-squared
amplitude of $120~\text{V}$ and a frequency of $60~\text{Hz}$ to emulate the
power supply of a typical household.
The impedance is chosen to be $Z_1(s) = R_1 + L_1 s$, where $R_1$ and $L_1$ are
determined using the geometry of the complex number that is the nominal grid
impedance. This impedance depends on the grid voltage $V_\text{grid}$, the DUT's
nominal power rating, the grid's short-circuit ratio $S$, and the grid's
inductance-to-resistance ratio.
Finally, as a current source cannot be causally connected in series with an
inductor, a shunt resistor $R_j$ of $1000~\Omega$ is inserted in parallel with
the source.

The DUT is chosen to be a series-interconnected pair of \textit{Dyna-Glo}
heaters on each phase, shown in Figure~\ref{fig:5_a_b}, the physics of each
being modeled as a resistive load \(R=12~\Omega\), yielding the circuit
depicted in Figure~\ref{fig:1_c} with \(Z_d(s) = R+R\).  The heaters are
actuated by an \textit{OP8110} power amplifier, depicted in
Figure~\ref{fig:5_a_b}, which is modeled by a continuous-time transfer function,
\begin{align} 
    A(s) = \frac{6.221\cdot 10^9}{s^2 + 1.255\cdot 10^5 s + 6.099\cdot 10^9},
\end{align}
that has been obtained experimentally using standard frequency-domain system identification techniques. The power amplifier is fed by an \textit{ITECH}
programmable direct current (DC) source.
% The DUT is chosen to be a resistive load, modeled as the circuit depicted in
% Figure~\ref{fig:1_c}, with $Z_d(s) = R_2$, where $R_2 = 24~\Omega$ represents a
% group of \textit{Dyna-Glo} heaters, shown in Figure~\ref{fig:5_b}. The heaters
% were actuated by an \textit{OP8110} power amplifier, depicted in
% Figure~\ref{fig:5_a}, whose experimentally identified continuous-time transfer
% function is
% \begin{align}
%     A(s) = \frac{6.221\cdot 10^9}{s^2 + 1.255\cdot 10^5 s + 6.099\cdot 10^9}.
% \end{align}
% The power amplifier is fed by an \textit{ITECH} programmable direct current (DC)
% source.

\begin{figure}[t]
    \centering
    \begin{tikzpicture}[scale=0.85]

\node [
    draw,
    rectangle,
    minimum height=15 mm,
    minimum width=10 mm,
] (ros) at (-3, 0) {ROS};
\node [
    draw,
    rectangle,
    minimum height=15 mm,
    minimum width=10 mm,
] (k) at (0, 0) {$\mbf{K}$};
\node [
    draw,
    rectangle,
    minimum height=15 mm,
    minimum width=10 mm,
] (dut) at (3, 0) {DUT};
\node [
    draw,
    rectangle,
    minimum size=6 mm,
] (d1) at (-1.5, 0.45) {$z^{-1}$};
\node [
    draw,
    rectangle,
    minimum size=6 mm,
] (d2) at (1.5, 0.45) {$z^{-1}$};
\node [
    draw,
    rectangle,
    minimum size=6 mm,
] (d3) at (1.5, -0.45) {$z^{-1}$};

\draw [<-] ($(ros.east) + (0, 0.45)$) -- (d1.west);
\draw [<-] (d1.east) -- ($(k.west) + (0, 0.45)$);
\draw [<-] ($(k.west) + (0, -0.45)$) -- ($(ros.east) + (0, -0.45)$);
\draw [->] ($(k.east) + (0, 0.45)$) -- (d2.west);
\draw [->] (d2.east) -- ($(dut.west) + (0, 0.45)$);
\draw [<-] ($(k.east) + (0, -0.45)$) -- (d3.west);
\draw [<-] (d3.east) -- ($(dut.west) + (0, -0.45)$);

\end{tikzpicture}
    \caption{Illustration of the delay structure inherent to the experiment
    reported in this paper. The incoming and outgoing channels of $\mbf{K}$ are
    the measurement and actuation channels, respectively. The delay in the ROS
    actuation channel is inserted to prevent algebraic-loop issues in the
    real-time simulation.}
    \label{fig:4}
\end{figure}
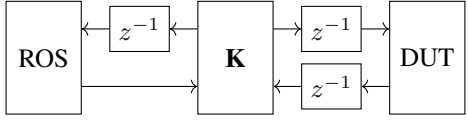

The interface to be designed, $\mbf{K}$, is a discrete-time LTI dynamical system
operated alongside the ROS circuit in \textit{Simulink}. This means that the
delay structure inherent to the experiment corresponds to that depicted in
Figure~\ref{fig:4}. These delays are caused by the communication conversions
between digital and analog signals present in the \textit{Simulink} model and
DUT, respectively. 

\begin{table}[t!]
    \centering
    \begin{tabular}{cc|cc}
        \textbf{Channel} & \textbf{Scaling} & \textbf{Channel} & \textbf{Scaling} \\
        \hline
        $\mbf{w} = V_\text{grid}$ & $120\sqrt{2}$~\unit{\volt} & ${z}_5 = V$ &
        $(200~\unit{\volt})^{-1}$ \\
        ${u}_1 = V$ & $200$~\unit{\volt} & ${z}_6 = J_B$ &
        $(15~\unit{\ampere})^{-1}$ \\
        ${u}_2 = J_B$ & $15$~\unit{\ampere} & $y_1 = V_1$ &
        $(120~\unit{\volt})^{-1}$ \\
        $z_1 = V_1 - V_\text{ref}$ & $(6~\unit{\volt})^{-1}$ & $y_2 = V_c$ &
        $(120~\unit{\volt})^{-1}$ \\
        $z_2 = I_1 - I_\text{ref}$ & $(0.5~\unit{\ampere})^{-1}$ & $y_3 = I_1$ &
        $(10~\unit{\ampere})^{-1}$ \\
        $z_3 = V_c - V_\text{ref}$ & $(6~\unit{\volt})^{-1}$ & $y_4 = I_d$ &
        $(10~\unit{\ampere})^{-1}$ \\
        $z_4 = I_d - I_\text{ref}$ & $(0.5~\unit{\ampere})^{-1}$ & &
    \end{tabular}
    \caption{Generalized plant scalings used in the experiment. The scalings on
    the signals of $\mbf{z}$ encode the performance specifications.}
    \label{tab:example}
\end{table}

\begin{figure}[]
    \centering
    \includegraphics[width=\columnwidth]{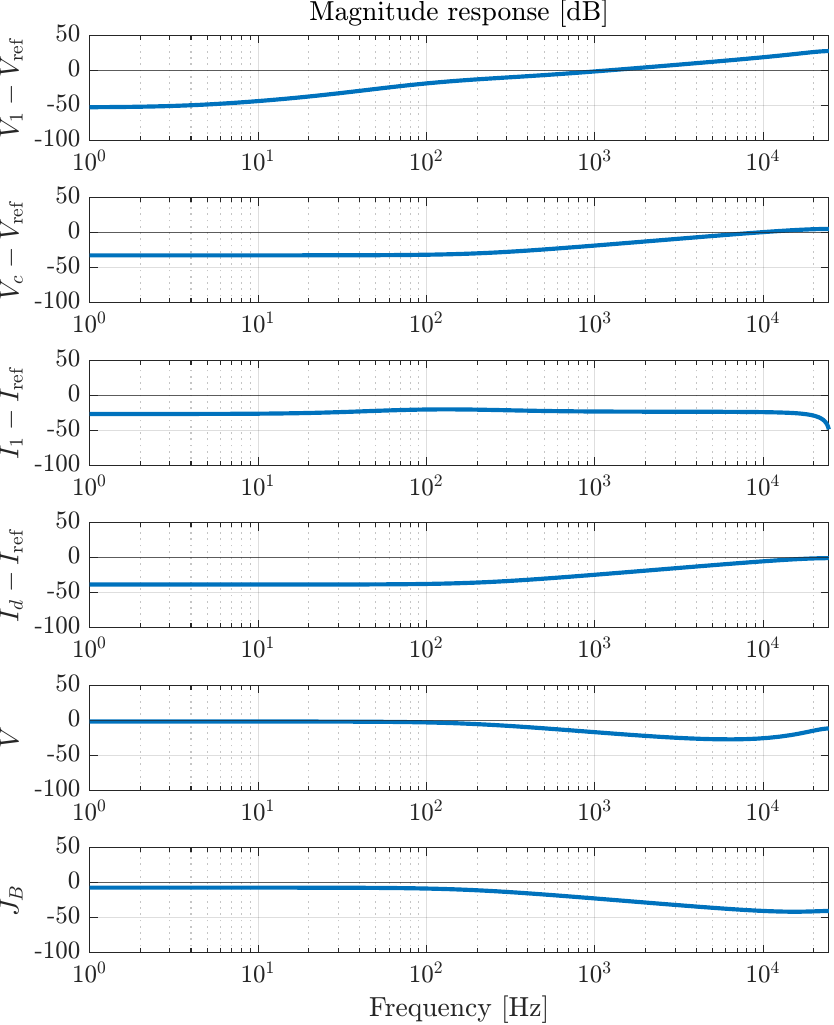}
    \caption{Frequency-domain magnitude response of the normalized closed loop.
    All component transfer functions have a gain less than $0$~dB in the $1$-kHz
    bandwidth, which guarantees that the steady-state time-domain response of the
    closed loop will be bounded within the prescribed specifications in that
    bandwidth.}
    \label{fig:freqresp}
\end{figure}

Each channel of the generalized plant is normalized by a scaling factor chosen
to reflect the maximum expected or allowable magnitude of that signal, as
outlined in Table~\ref{tab:example}.

% \begin{table}[t!]
%     \centering
%     \begin{tabular}{cc|cc}
%         \textbf{Channel} & \textbf{Scaling} & \textbf{Channel} & \textbf{Scaling} \\
%         \hline
%         $\mbf{w} = V_\text{grid}$ & $120\sqrt{2}$~\unit{\volt} & ${z}_5 = V$ &
%         $(200~\unit{\volt})^{-1}$ \\
%         ${u}_1 = V$ & $200$~\unit{\volt} & ${z}_6 = J_B$ &
%         $(15~\unit{\ampere})^{-1}$ \\
%         ${u}_2 = J_B$ & $15$~\unit{\ampere} & $y_1 = V_1$ &
%         $(120~\unit{\volt})^{-1}$ \\
%         $z_1 = V_1 - V_\text{ref}$ & $(6~\unit{\volt})^{-1}$ & $y_2 = V_c$ &
%         $(120~\unit{\volt})^{-1}$ \\
%         $z_2 = I_1 - I_\text{ref}$ & $(0.5~\unit{\ampere})^{-1}$ & $y_3 = I_1$ &
%         $(10~\unit{\ampere})^{-1}$ \\
%         $z_3 = V_c - V_\text{ref}$ & $(6~\unit{\volt})^{-1}$ & $y_4 = I_d$ &
%         $(10~\unit{\ampere})^{-1}$ \\
%         $z_4 = I_d - I_\text{ref}$ & $(0.5~\unit{\ampere})^{-1}$ & &
%     \end{tabular}
%     \caption{Generalized plant scalings used in the experiment. The scalings on
%     the signals of $\mbf{z}$ encode the performance specifications.}
%     \label{tab:example}
% \end{table}

For the shaping of the closed-loop dynamics, bilinear-transform discretizations
of continuous-time filters are used. The exogenous signal $\mbf{w}(z) =
V_\text{grid}(z)$ is filtered with a low-pass filter with a cutoff of
$1~\unit{\kilo\hertz}$ to encode its nominal frequency of $60~\unit{\hertz}$ and
possible higher-frequency modes. The error signals, $z_1(z)$ through $z_4(z)$ in
objective~(\ref{eqn:transparency_objective}), are filtered with similar low-pass
filters to minimize them in the frequency bandwidth of $\mbf{w}(z)$. The
actuation signals, $z_5(z)$ and $z_6(z)$ in
objective~(\ref{eqn:transparency_objective}), are filtered with high-pass
filters with cutoffs of $1~\unit{\kilo\hertz}$ to allow for sufficient control
effort in the bandwidth of $\mbf{w}(z)$, and minimize it in the noise bandwidth.
The result of this loop shaping is depicted in Figure~\ref{fig:freqresp}.

The synthesized interface is then ready for insertion into the \textit{Simulink}
model, and for the closed-loop, real-time actuation of the PHIL experiment
components.

The interface algorithm synthesized using the proposed method is then compared
against an ITM-based interface. Similar to what was briefly described in
Section~I, the ITM-based interface actuates the DUT voltage source $V$ with the
ROS voltage $V_1$, and the ROS current source $J_B$ with a filtered version of
the DUT current $I_d$. The filter parametrizes the tradeoff between the
interface's range of stability and its accuracy.

For the purpose of this experiment, the stability range of an interface is
parametrized by the ROS short-circuit ratio $S$. The ITM-based interface
is found to maintain stability and acceptable levels of accuracy at
short-circuit ratios $S \geq 5$, with accuracy degrading drastically at lower
values of $S$. As is expected of ITM-based
interfaces~\cite{tremblay1}, the interface could not maintain
stability below a threshold short-circuit ratio $S^*$, which was experimentally
deduced to satisfy $1 < S^* \leq 2$ for the particular filter used. Given this
information, the model matching-based interface is designed with $S = 1$ as the
nominal grid operating condition. This interface is then operated in various
short-circuit conditions, $S \in \{0.1, 1, 2, 5, 200\}$, and found to
maintain stability and acceptable levels of accuracy for all of them.

\begin{figure}[t]
    \centering
    % \begin{subfigure}[b]{0.49\columnwidth}
    %     \centering
    %     \includegraphics[
    %         width=0.94\columnwidth
    %     ]{./figures/test_bench_.jpeg}
    %     \caption{OPAL-RT PHIL test bench with the simulator (top), power
    %     amplifiers (center), and programmable DC source (bottom).\\}
    %     \label{fig:5_a}
    % \end{subfigure}
    % \begin{subfigure}[b]{0.49\columnwidth}
    %     \centering
    %     \includegraphics[width=0.94\columnwidth]{./figures/test_load_.jpeg}
    %     \caption{Group of heaters as the DUT.\\}
    %     \label{fig:5_b}
    % \end{subfigure}
    % \hfill
    % \begin{subfigure}[b]{\columnwidth}
        \includegraphics[width=\linewidth]{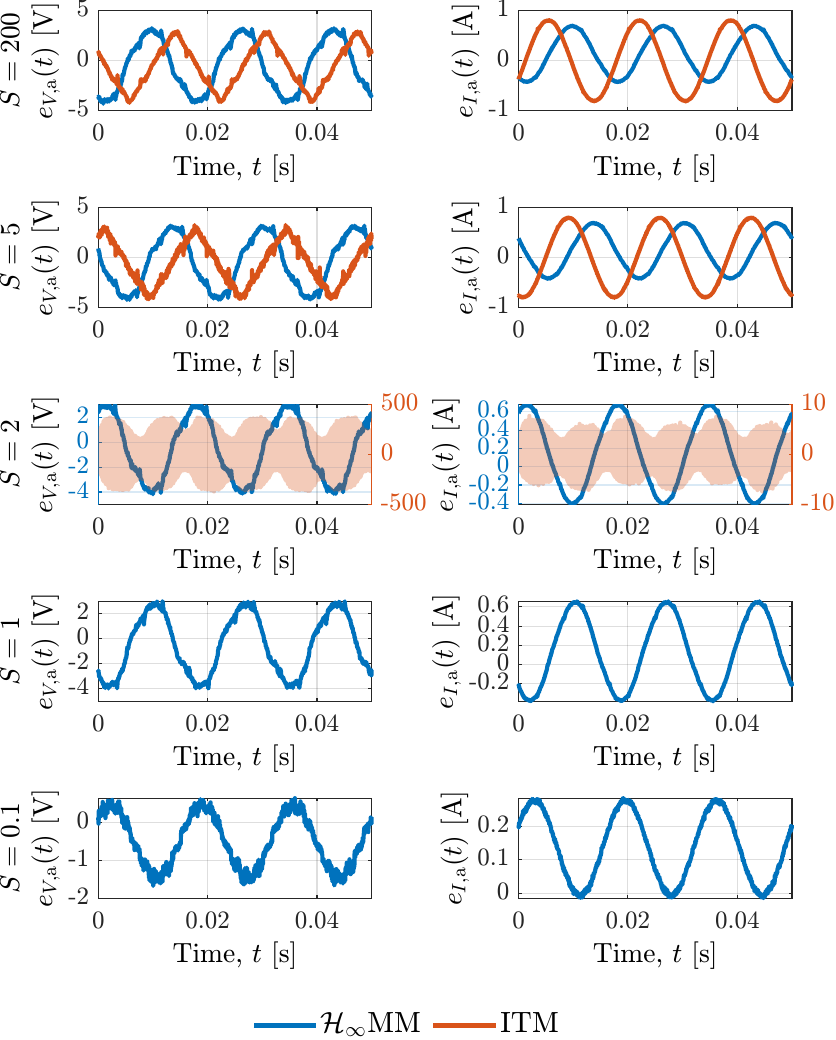}
        \caption{
            Time-domain responses of the model matching- and ITM-based
            interfaces under various short-circuit conditions $S$. The
            left column shows the accuracy in voltage matching of phase A,
            $e_{V,\mathrm{a}}(t) = V_{1,\mathrm{a}}(t) - V_{c,\mathrm{a}}(t)$.
            The right column shows the accuracy in current matching of phase A,
            $e_{I,\mathrm{a}}(t) = I_{1,\mathrm{a}}(t) - I_{d,\mathrm{a}}(t)$.
            When $S \in \{200, 5\}$, the two interfacing approaches exhibit
            comparable accuracy. When $S = 2$, the ITM-based interface sees its
            performance drastically degraded, while the model matching-based
            method does not. When $S \in \{1, 0.1\}$, the model matching-based
            interface continues to exhibit good accuracy, while the ITM-based
            interface is unstable, and therefore not shown.
        }
        \label{fig:5_c}
    % \end{subfigure}
    % \caption{PHIL experimental results.}
    % \label{fig:5}
\end{figure}

Figure~\ref{fig:5_c} depicts an accuracy comparison of the model matching-based
and ITM-based interfaces in the various short-circuit conditions. The
model matching-based interface's accuracy, obtained by designing for
\textit{transparency}, is comparable to that of the ITM-based interface in
off-nominal operating conditions of high short-circuit ratio. The model
matching-based approach drastically outperforms the ITM-based one in close-to-nominal
conditions, where $S = 2$. And finally, in off-nominal conditions of low
short-circuit ratio, the model matching-based interface is able to stabilize the
closed loop with good accuracy when the ITM-based one is not.

These experimental results demonstrate the viability of the proposed model
matching-based interfacing approach as an improvement to the ITM-based one,
especially when employed around the ITM-based interface's short-circuit ratio
threshold of instability.

\section{Conclusion}\label{sec:5}

The paper proposes a model matching-based approach to the design of a PHIL
interfacing algorithm as an alternative to the ITM-based and existing
$\mathcal{H}_\infty$ control approaches. The model matching-based approach
reformulates the standard $\mathcal{H}_\infty$ control objective of accuracy in
favor of the more general objective of transparency.

The model matching-based methodology is experimentally shown to be viable for
use in real-time PHIL setups, and is reported to achieve accuracy levels and a
stability range that are comparable or superior to those of the ITM-based
methodology.

Finally, the model matching-based interface tested in the experiment is found to
exhibit very good robustness. However, it should be remarked that no
considerations of robustness are made in the model matching-based design
methodology. The reported improvement in the stability range therefore cannot be
formally attributed to the model matching-based approach, but will serve as an
important starting point for future investigation into the design constraint of
robustness.

\printbibliography

@article{bhattarai,
title = {Assay of renewable energy transition: A systematic literature review},
journal = {Sci. Total Environ.},
volume = {833},
pages = {155159},
year = {2022},
author = {Utsav Bhattarai and Tek Maraseni and Armando Apan}
}

@Article{vonjouanne,
AUTHOR = {von Jouanne, Annette and Agamloh, Emmanuel and Yokochi, Alex},
TITLE = {Power Hardware-in-the-Loop ({PHIL}): A Review to Advance Smart Inverter-Based Grid-Edge Solutions},
JOURNAL = {Energies},
VOLUME = {16},
YEAR = {2023},
NUMBER = {2},
ARTICLE-NUMBER = {916},
}

@ARTICLE{lauss,
  author={Lauss, Georg F. and Faruque, M. Omar and Schoder, Karl and Dufour, Christian and Viehweider, Alexander and Langston, James},
  journal={{IEEE} Trans. Ind. Electron.}, 
  title={Characteristics and Design of Power Hardware-in-the-Loop Simulations for Electrical Power Systems}, 
  year={2016},
  volume={63},
  number={1},
  pages={406-417}
}

@INPROCEEDINGS{ren,
  author={Ren, W. and Steurer, M. and Baldwin, T. L.},
  booktitle={{2007 IEEE/IAS Ind. Commercial Power Syst. Tech. Conf.}}, 
  title={Improve the Stability and the Accuracy of Power Hardware-in-the-Loop Simulation by Selecting Appropriate Interface Algorithms}, 
  year={2007},
  volume={},
  number={},
  pages={1-7}
  }

@INPROCEEDINGS{naerum,
  author={Naerum, Edvard and Hannaford, Blake},
  booktitle={{2009 IEEE Int. Conf. Robot. Autom.}}, 
  title={{Global transparency analysis of the Lawrence teleoperator architecture}}, 
  year={2009},
  volume={},
  number={},
  pages={4344-4349}
  }

@Article{resch,
AUTHOR = {Resch, Simon and Friedrich, Juliane and Wagner, Timo and Mehlmann, Gert and Luther, Matthias},
TITLE = {Stability Analysis of Power Hardware-in-the-Loop Simulations for Grid Applications},
JOURNAL = {Electronics},
VOLUME = {11},
YEAR = {2022},
NUMBER = {1},
ARTICLE-NUMBER = {7},
}

@ARTICLE{salapaka1,
  author={Lundstrom, Blake and Salapaka, Murti V.},
  journal={IEEE Trans. Ind. Electron.}, 
  title={Optimal Power Hardware-in-the-Loop Interfacing: Applying Modern Control for Design and Verification of High-Accuracy Interfaces}, 
  year={2021},
  volume={68},
  number={11},
  pages={10388-10399}
}

@ARTICLE{salapaka2,
  author={Chakraborty, Soham and Rana, Mohammed Tuhin and Salapaka, Murti V.},
  journal={IEEE Trans. Ind. Electron.}, 
  title={A Robust Power Hardware-in-the-Loop Interface Under Uncertain Software and Hardware System}, 
  year={2025},
  volume={72},
  number={6},
  pages={6088-6102}
}

@book{scherer,
  author    = {Scherer, Carsten},
  title     = {{Theory of Robust Control}},
  url       = {https://www.imng.uni-stuttgart.de/mst/files/RC.pdf},
}

@book{skogestad,
author = {Skogestad, Sigurd and Postlethwaite, Ian},
title = {{Multivariable Feedback Control: Analysis and Design}},
year = {2005},
publisher = {John Wiley \& Sons, Inc.},
address = {Hoboken, NJ, USA}
}

@book{zhou_doyle_glover,
author = {Zhou, Kemin and Doyle, John C. and Glover, Keith},
title = {{Robust and Optimal Control}},
year = {1996},
publisher = {Prentice-Hall, Inc.},
address = {Englewood Cliffs, NJ, USA}
}

@article{deoliveira_geromel_bernussou,
author = {M. C. De Oliveira and J. C. Geromel and J. Bernussou},
title = {Extended $\mathcal{H}_2$ and $\mathcal{H}_\infty$ norm characterizations and controller parametrizations for discrete-time systems},
journal = {International Journal of Control},
volume = {75},
number = {9},
pages = {666--679},
year = {2002},
publisher = {Taylor \& Francis},}

@misc{caverly_forbes,
      title={{LMI Properties and Applications in Systems, Stability, and Control Theory}}, 
      author={Ryan James Caverly and James Richard Forbes},
      year={2024},
      eprint={1903.08599},
      archivePrefix={arXiv},
      primaryClass={eess.SY},
      url={https://arxiv.org/abs/1903.08599}, 
}

@article{tremblay1,
author = {Tremblay, Olivier and Fortin-Blanchette, Handy and Gagnon, Richard and Brissette, Yves},
title = {Contribution to stability analysis of power hardware-in-the-loop simulators},
journal = {IET Generation, Transmission \& Distribution},
volume = {11},
number = {12},
pages = {3073-3079},
year = {2017}
}

@ARTICLE{tremblay2,
  author={Tremblay, Olivier and Rimorov, Dmitry and Gagnon, Richard and Fortin-Blanchette, Handy},
  journal={IEEE Transactions on Energy Conversion}, 
  title={A Multi-Time-Step Transmission Line Interface for Power Hardware-in-the-Loop Simulators}, 
  year={2020},
  volume={35},
  number={1},
  pages={539-548}}

\end{document}